\documentclass[twocolumn,natbib209]{emulateapj}    

\begin{document}
\title{A distinct peak-flux distribution of the third class of\\
gamma-ray bursts: A possible signature of X-ray flashes?}

\author{P. Veres  \altaffilmark{}}
\affil{Department of Physics of Complex Systems, E\"otv\"os University,
H-1117 Budapest, P\'azm\'any P. s. 1/A, Hungary \\ 
Department of Physics, Bolyai Military University, H-1581
Budapest, POB 15, Hungary}
\email{veresp@elte.hu}

\author{Z. Bagoly \altaffilmark{}}
\affil{Dept. of Physics of Complex Systems, E\"otv\"os University,
H-1117 Budapest, P\'azm\'any P. s. 1/A, Hungary}

\author{I. Horv\'ath \altaffilmark{}}
\affil{Department of Physics, Bolyai Military University, H-1581
Budapest, POB 15, Hungary}

\author{A. M\'esz\'aros \altaffilmark{}}
\affil{Charles University, Faculty of Mathematics and Physics,
        Astronomical Institute,  V Hole\v{s}ovi\v{c}k\'ach 2, 180 00  
		Prague 8,      Czech Republic }

\author{L. G. Bal\'azs  \altaffilmark{}}
\affil{Konkoly Observatory, H-1505 Budapest, POB 67, Hungary}
\begin{abstract}
Gamma-ray bursts are the most luminous events in the Universe. Going
beyond the short-long classification scheme we work in the context of three
burst populations with the third group of intermediate duration and softest
spectrum.  We are looking for physical properties which discriminate the
intermediate duration bursts from the other two classes.  We use maximum
likelihood fits to establish group memberships in the duration-hardness plane.
To confirm these results we also use k-means and hierarchical clustering.  We
use Monte-Carlo simulations to test the significance of the existence of the
intermediate group and we find it with $99.8\%$ probability.  The intermediate
duration population has a significantly lower peak-flux (with $99.94\%$
significance). Also, long bursts with measured redshift have  higher peak-fluxes
(with $98.6\%$ significance) than long bursts without measured redshifts.  As
the third group is the softest, we argue that we have { related} them with
X-ray flashes among the gamma-ray bursts.  We give a new, probabilistic
definition for this class of events.  
\end{abstract}


\maketitle

\section{Introduction}
Gamma-ray bursts (GRBs) are the most powerful explosions known in the Universe
(for a review see \citet{2006RPPh...69.2259M}).  To discern the physical
properties of GRBs as a whole, we need to understand the number of physically
different underlying classes of the phenomenon
\citep{2009ApJ...703.1696Z,2010arXiv1001.0598L}.

Before the launch of BATSE \citep{1994ApJS...92..229F}, there were hints of two
distinct populations \citep{maz81,nor84}.  The bimodality was established using
BATSE observations of the duration by \citet{kou93}. The subsequent classes
were dubbed short and long type GRBs referring to their durations.  More
sophisticated statistical methods based on more data { using one
classification parameter \citep{hor98} }  and more than one observable
property, showed three populations in the BATSE data \citep{muk98}.  These were
further confirmed by subsequent analyses \citep{hak03,hor06,chat07}: the third
population is intermediate in duration \citep{hor02}. 

Many statistical methods (e.g., maximum likelihood fitting, chi-square fitting,
clustering) point to the presence of the third class with high significance.
These methods reveal three groups in the data from different satellites (BATSE
\citep{hor06}, BeppoSAX \citep{2009Ap&SS.323...83H}, RHESSI
\citep{2009A&A...498..399R}, Swift \citep{hor08,2009A&A...504...67H,hor10}).
These independent observations  show that there is a good reason for the
reality of the intermediate population. 

While the existence of the intermediate population is proven with high
significance using data from different experiments and using different
statistical methods, a physical model to explain the origin of this third
population is still missing \citep{2006RPPh...69.2259M}.

With the launch of the Swift satellite \citep{2004ApJ...611.1005G}, a new
perspective has opened up of the study of gamma-ray bursts and their
afterglows.  The intermediate population in the studies so far has always been
the softest among the groups, meaning intermediate GRBs emit the bulk of their
energy in the low-energy gamma-rays. Swift's gamma-ray detector BAT has an
energy coverage from $15$ to $150$ keV - hence Swift is well suited for the
study of the intermediate population. 

Here we report on a significant difference in the peak-flux distribution
between the intermediate and the short and between the intermediate and long
populations. We identify a third population using a multi-component model and
we show that this group { has a significant overlap} with  X-ray flashes. We
give a probabilistic definition of this class.

We present our sample in Section 2. In the next section we perform the
classification with three methods and discuss the stability of the clustering.
In Section 4. we present the peak-flux distribution of the classes. In Section
5. we analyze the samples with and without measured redshift. In Section 6.  we
interpret the findings and in Section 7. we conclude by summarizing the paper's
results.


\section{Sample}
The First Swift BAT Catalog \citep{2008ApJS..175..179S} was augmented with
bursts up to August 7, 2009 with measured $T_{90}$ and hardness ratio.  After
excluding the outliers and bursts without measured parameters, the sample
consisting of the \citet{2008ApJS..175..179S} sample and our extension has a
total of $408$ GRBs ($219$ from the Catalog and $189$ newer bursts). 

Data reduction was carried out using HEAsoft version 6.3.3 and calibration
database version 20070924. For light curves and spectra we ran the
\texttt{batgrbproduct}\footnote{http://heasarc.nasa.gov/lheasoft/} pipeline.  To
obtain the spectral parameters we fitted the spectra integrated for the
duration of the burst with a power law model and a power law model with an
exponential cutoff. As in \citet{2008ApJS..175..179S} we have chosen the cutoff
power law model if the $\chi^2$ of the fit improved by more than $6$. 

The most widely used duration measure is $T_{90}$, which is defined as the
period between the $5\%$ and $95\%$ of the incoming counts.  To find the
fluences ($S_{E_{min},E_{max}}$) we integrated the model spectrum in the usual
Swift energy bands with $15-25-50-100-150$ keV as their boundaries.  We define
the hardness ratio ($H_{ij}$, where $i$ and $j$ mark the two energy intervals)
as the ratio of the fluences in different channels for a given burst.  For
example $H_{32} = \frac{S_{50-100}}{S_{25-50}}$, where $S_{50-100}$ is the
fluence of the burst for the entire duration measured in the $50-100$ keV
range.  Different hardness ratios are possible to define and we have used
them to check our results.  

Bursts have a wealth of measured parameters and it is possible to use many
variations of them. The choice of $T_{90}$ has some draw-backs {
\citep{2010ScChG.tmp...34Q}}. It is not sensitive to quiescent episodes between
the active phase of bursts (e.g., bursts with precursors). Also it cannot
differentiate between bursts with an initial hard peak and a soft extended
emission from bursts with constant long emission.  In turn this latter type of
burst with a hard initial spike and an extended soft emission can bias $H_{32}$
as well \citep{2006Natur.444.1044G}.  Nevertheless, keeping in mind these
draw-backs of $T_{90}$, this quantity is still one of the most important
measures of GRBs, and hence its use is straightforward. This question is
discussed also in Section \ref{robust}.

\section{Classification}
\subsection{The choice of variables}
There are many indications that the phenomenon which we observe as gamma-ray
bursts has more than one underlying population. The goal is to identify classes
which are physically different. We choose the duration and the hardness ratio
of bursts as the principal measure. This choice has been made by other studies
as well \citep{1996ApJ...471L..27D,hor06}. 

The choice of variables for the clustering deserves some justification.
Satellites generally observe many properties and subsequent observations add to
the volume of the parameters belonging to a burst.  \cite{bag98} showed that
two principal components are enough to describe the data in the BATSE 
Catalog satisfactorily.  \cite{hor06} followed these arguments and used the
$H_{32}$ hardness and the $T_{90}$ duration to classify the bursts. By using
$T_{90}$ and $H_{32}$ we include a basic temporal and spectral characteristic
of the bursts.

The reality of any classification is hard to assess.  A good way to make sure
that the classification is robust and has some physical significance is to
check the groups' stability with respect to various classification methods.  We
carry out three types of classifications: model-based multivariate
classification, k-means clustering and hierarchical clustering.

In the mathematical literature there is a wealth { of} classification
schemes. We use { the algorithms implemented} in the R
software\footnote{http://cran.r-project.org \citep{R}}. The clustering methods
can be divided to parametric and non-parametric schemes. Parametric schemes
postulate that the data follows a pre-defined model (in our case a
superposition of multi-variate Gaussian distributions) and give a membership
probability for each  gamma-ray burst belonging to a given group. Thus each
burst will have assigned $k$ number of { membership} probabilities, where
$k$ is the number of multivariate components (groups).  This is called a fuzzy
clustering \citep{Yang19931}. The non-parametric tests (k-means and
hierarchical clustering), on the contrary, assign definitive memberships to
each burst.  { However, here one needs to define the distance or similarity
measure between the cases.} 

\subsection{Model based clustering}\label{modfit}
As discussed in \cite{hor06}, we can assume that the observed distribution of
bursts on the duration-hardness plane is a superposition of two or more groups.
The conditional probability density ($p(\log_{10} T_{90}, \log_{10} H_{32}|l)
$), together with the probability of a burst being from a given group ($p_l$)
using the law of full probabilities: 

\begin{equation}\label{eq1}
p(\log_{10} T_{90}, \log_{10} H_{32}) =
\sum_{l=1}^k p(\log_{10} T_{90}, \log_{10} H_{32}|l) p_l, 
\end{equation}
where $k$ is the number of groups.

Studies show that for example the distribution of the logarithm of the duration
can be adequately described by a superposition  of three Gaussians
\citep{hor98}.  In this section we thus use the model based on bivariate
Gaussian distributions.  We suppose that the joint distribution of the
parameters can be described as a superposition of Gaussians.  Previously
\cite{hor06} carried out a similar analysis on the duration-hardness plane of
the BATSE Catalog where data was fitted with bivariate Gaussians.

One bivariate Gaussian will have the following joint distribution function:

\begin{widetext}
\begin{eqnarray} 
p(\log_{10} T_{90},\log_{10} H_{32}|l) = \frac{1}{2 \pi \sigma_{\log_{10}
T_{90}} \sigma_{\log_{10} H_{32}} \sqrt{1-r^2}} \exp \left( -
\frac{1}{2(1-r^2)} \left(\frac{(\log_{10} T_{90}-{\log_{10}
T_{90}}_C)^2}{\sigma_{\log_{10} T_{90}}^2}+ \right. \right. \nonumber \\
\left. \left. +\frac{(\log_{10} H_{32}-{\log_{10}
H_{32}}_C)^2}{\sigma_{\log_{10} H_{32}}^2} + \frac{2 r (\log_{10}
T_{90}-{\log_{10} T_{90}}_C)(\log_{10} H_{32}-{\log_{10}
H_{32}}_C)}{\sigma_{\log_{10} T_{90}}\sigma_{\log_{10} H_{32}}}  \right)
\right),
\end{eqnarray}
\end{widetext}
where ${\log_{10} T_{90}}_C$ and ${\log_{10} H_{32}}_C$ are the ellipse center
coordinates, $\sigma_{\log_{10} T_{90}}$ and $\sigma_{\log_{10} H_{32}}$ are
the two standard deviations of the distribution and $r$ is the correlation
coefficient. 

Here we find the model parameters using the maximum likelihood method. The
procedure is called Expectation-Maximization (EM). This consists of appointing
a membership probability to each burst using an initial value of the parameters
(E step). Then we calculate the parameters of the model using these memberships
{(M step)}.  Using this new model we re-associate each burst to the groups
and calculate the model parameters. We repeat these steps until the solution
converge.  It is proved that this procedure converges to the maximum likelihood
solution of the parameters \citep{dempster77em}.

\subsection{Number of groups}\label{vei}
It is important to decide on the { true}  number of components to fit (the
number of classes). In the model-based framework we have a better grip on this
problem compared to the non-parametric methods.  For our calculations we use
the Mclust package\footnote{http://www.stat.washington.edu/mclust}
\citep{Fraley00model-basedclustering} of R.  

In the most general case the best model is found by maximizing the likelihood.
It is possible to penalize a model for more degrees of freedom.  A widely used
version of this method is called the Bayesian Information Criterion (BIC)
introduced by \cite{ISI:A1978EQ63300014} (for astronomical applications see
e.g.  \citet{2007MNRAS.377L..74L}).  The function to be maximized to get the
best fitting model parameters has an additional term besides the
log-likelihood:
\begin{equation} BIC = 2  L_{max}  - m \ln N, \end{equation} 
where $L_{max}$(=$\ln l_{max}$) is the logarithm of the maximum likelihood of
the model, $m$ is the number of free parameters, and $N$ is the size of the
sample. This method takes into account the complexity of our model by
penalizing for additional free parameters. 

We use the BIC to find the most probable model (including the number of
components) and the parameters of this model.  In a two-dimensional fit the
number of free parameters of a single bivariate Gaussian component is $6$ (two
coordinates for the mean, two values of the standard deviations in different
directions, a correlation coefficient and a weight). For $k$ bivariate
Gaussians the number of free parameters is $6k-1$, since the sum of the weights
is $1$.

In the most general model all $6$ parameters of each component can be varied.
Some of the parameters may have interrelations between the components (e.g.,
all components have the same weight or shape, there is no correlation between
the variables ($r=0$), etc.). In this way we construct models with less degrees
of freedom.  The possible interrelations between the parameters of the
Gaussians are taken into account by trying different models with different
types of constraints (for the list of models see the Mclust
manual\footnote{http://www.stat.washington.edu/research/reports/2006/tr504.pdf}).

\begin{figure}
\includegraphics[angle=-90,width=.5\textwidth]{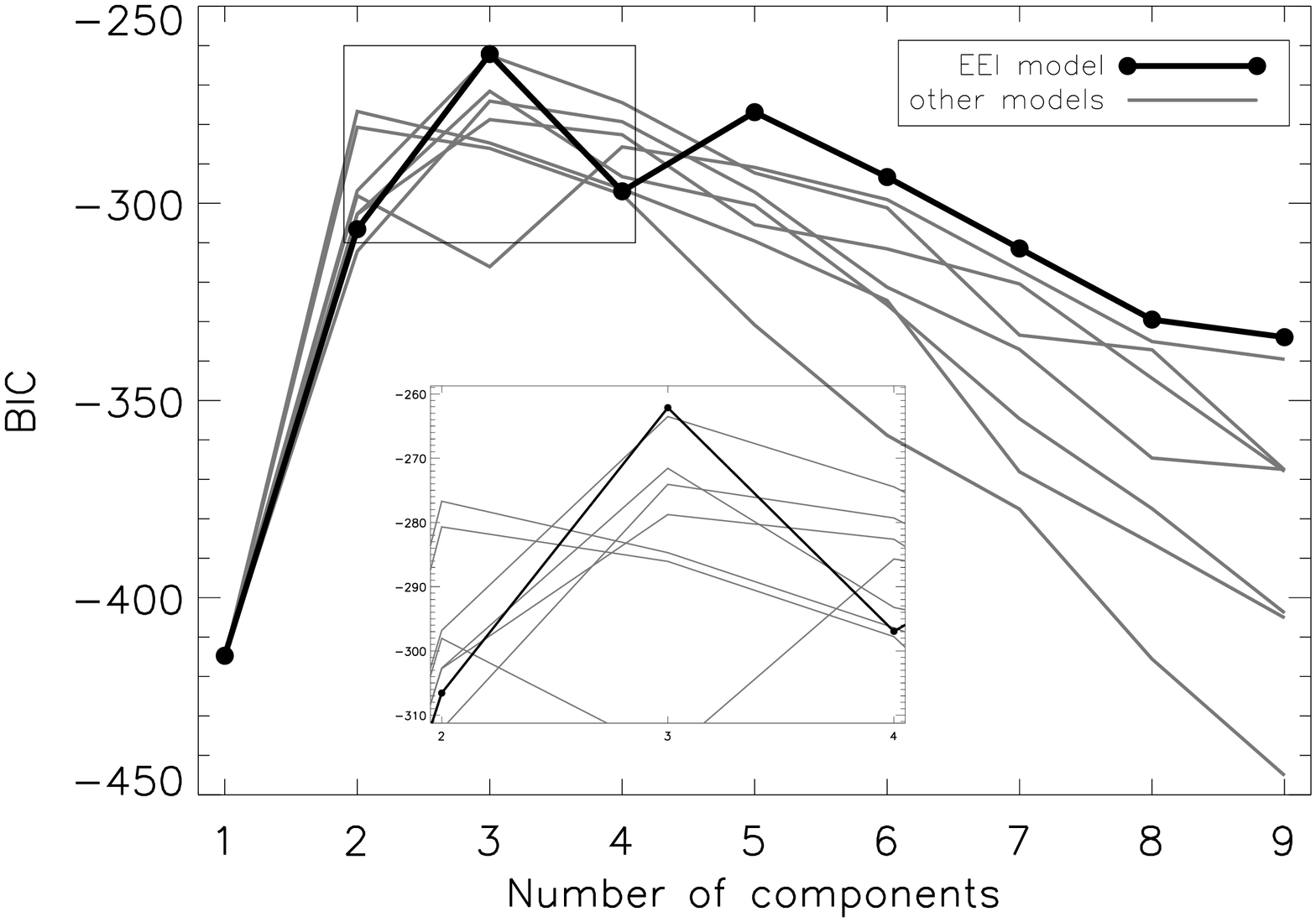} 
\caption{Bayesian information criterion values for different models 
(different lines show different models) in function of the number of bivariate
Gaussian components: the higher the value of the BIC, the more probable the
model. The best model is marked in black and the highest value is reached for
$k=3$. Also, some of the other models have their peak at $k=3$. The inset shows
the magnified peak region.\label{bic}} \end{figure}

We have applied this classification scheme on our sample.  We found that the
model { with three components gives the best fit for the data in the BIC
sense},  where the shape of the bivariate Gaussians is the same
($\sigma_{\log_{10} T_{90},i}=\sigma_{\log_{10} T_{90},j}$ and
$\sigma_{\log_{10} H_{32},i}=\sigma_{\log_{10} H_{32},j}$ for $i, j =$ \{short,
long, intermediate\}) for each group, only their weights are different with no
correlation. This is called the EEI model in Mclust. The description of the
model follows from its name: equal volume (E), equal shape (E) and the axes are
parallel with the coordinate axes (I). In other words this is the model with
optimal information content describing the data (see Fig. \ref{bic}).

We find that the best model has a value of $BIC=-262.14$. This model has three
bivariate components. { In the general case} the maximum number of free
parameters would thus be $m=17$.  Taking into account the constraints of this
model, the degrees of freedom here will be $m=10$ (three coordinate pairs for
the center of the distributions, two standard deviations common for all three
components and two weights). 

The clustering method based on this model shows that a model containing three
bivariate components is the most preferred. Models with two components have the 
best  $BIC \sim -276$ and for models with four components the best $BIC \sim
-274$, both are clearly below the maximum.  In case of a maximum likelihood
fit, we can infer the probability of the chance occurrence of a model compared
to another. In this case the difference in the value of the BIC of two models
informs us about the goodness of the model.  According to \cite{muk98} and
references therein, differences in BIC in the $8$-$10$ range represent strong
evidence in favor of the model with the higher BIC. In our case the differences
are even bigger.

The best-fit model has $10$ free parameters and has  three bivariate Gaussian
components.  The parameters of the model as well as the number of the bursts in
the groups are in Table \ref{modelparam}.  The shortest and hardest group will
be designated short, the longest and of moderate hardness will be called long
and the intermediate duration group with the softest spectrum will be called
intermediate. For the relation of the intermediate class to other studies on
this topic see the discussion.

\begin{table}  
\caption{Bivariate model parameters for the best-fitted (EEI) model. The standard
deviations in the direction of the coordinate axes and the correlation
coefficients are constrained by the model.\bigskip}
\label{modelparam}
  \begin{tabular}{cccccccc} \hline
  Groups 			& $p_l$ 	&${\lg T_{90}}_C$ 		& ${\lg
  H_{32}}_C$ 	& $\sigma_{\lg T_{90}}$& $\sigma_{\lg H_{32}}$ &
  $r$ & $N_l$ \\
 \hline \hline
short	          	& $0.08$   &   $-0.331$&    $ 0.247$ 	& $0.509$ 	& $0.090$	& $0$	& $31$	\\
interm.		    	& $0.12$	&   $ 1.136$&    $-0.116$	& $0.509$	& $0.090$	& $0$	& $46$	\\
long				& $0.80$  	&   $ 1.699$&    $ 0.114$ 	& $0.509$	& $0.090$	& $0$	& $331$	\\
\hline
\end{tabular}
\end{table}

To calculate the significance of three populations we carried out a Monte-Carlo
simulation.  We tested the hypothesis that the presence of the third population
is only a statistical fluctuation. We generated $10000$ random catalogs, with
the best $k=2$ model. We found that with the classification method only
$0.2\%$ of the cases yielded a three component model while $99.8\%$ of cases
produced a two component fit.  This means that the probability that the third
group is only a statistical fluctuation is $0.2\%$

To test the validity of the simulation{ ,} we have simulated again $10000$ samples
of $408$ bursts, using the model parameters for the three population model in
Table \ref{modelparam}. We found that two populations are statistical
fluctuation in $2.1\%$ of the cases compared to the three population model.

There is another three component model which has a very similar BIC value (the
difference is { only} $\sim 1$, see the inset in Fig. \ref{bic}). This model
is called VEI and it has variable volume, meaning the product of the standard
deviations is the same (V), equal shape (E) and the axes are parallel with the
coordinate axes (I). It has $12$ degrees of freedom (three coordinate pairs for
the centers of the distributions, three pairs of standard deviations with the
restriction that their product is the same (four degrees of freedom) and two
weights).  Information Criterion gives only a weak hint as to which model is
preferred.  The VEI model gives a visibly different group structure. If we
compare it to the three component EEI model the ratio of differently classified
bursts is $26.7\%$.  The group structure found by this model resembles the one
found by \citet{hor10} and will be addressed in the discussion section.

We assign class memberships using the ratio of the fitted bivariate models at
the burst location on the duration-hardness plane. We call these membership
probabilities. Any given burst is assigned to the group with the highest
probability.  Fuzzy classification \citep{Yang19931} implies that we can define
an indicator function in the following manner:
\begin{widetext}
\begin{equation}\label{pind}
I_\mathrm{l}(\log_{10} T_{90},\log_{10} H_{32})=\frac{P_\mathrm{l}\times
P(\log_{10} T_{90},\log_{10} H_{32}|l)}{\displaystyle\sum_{l
\in \mathrm{\{short, interm., long\}}} P_\mathrm{l}\times P(\log_{10}
T_{90},\log_{10} H_{32}|l)}.
\end{equation}
\end{widetext}
Here $P(\log_{10} T_{90},\log_{10} H_{32}|l)$ is the conditional probability
density of a burst, assuming it comes from class $l$. $P_l$ is the probability
of the $l$ class. The indicator function assigns a probability for a burst that
it belongs to a given group.\footnote{for the detailed classification results
with the EEI model see: http://itl7.elte.hu/$\sim$veresp/swt90h32gr408.txt}

In this framework there is no definite answer to the question: {\it "To which
class does a specific burst belong?"}, rather there is a probability of a burst
belonging to a given group as given by the indicator function. If the
contribution of a component is dominant, the membership determination is
straightforward. If the two (or three) highest membership probabilities for a
burst are approximately equal the uncertainty in the classification is high.
To check for contamination from other groups we carry out our analysis on a
sub-sample where only the more certain memberships are taken into
consideration. 

The model has three components with equal standard deviations in both
directions and with no correlation ($r=0$). The components are as follows (see
also Table \ref{modelparam}): 
\begin{enumerate}

\item The first component is the known {\it short} class of GRBs (shortest
duration and hardest spectra). The average duration is $0.47$ s and the average
hardness ratio is $1.77$. It has $31$ members, and the weight of this model
component is  $0.08$.

\item The second, most numerous model component is the the {\it long} class,
also identified in many previous studies. It has an average duration of $50.0$
s and an average hardness of $1.30$. It has $331$ members and the weight of the
model component is $0.80$

\item The third and softest class is {\it intermediate} in duration. It {
has overlapping regions}  with previous definitions of the intermediate class
{ \citep{hor10}}.  The average duration is $13.7$ s, and the average hardness of
this class is $0.77$.  It has $46$ members and the weight of the model
component is $0.12$.  
\end{enumerate}

All components have the same standard deviation in both directions. This means
the shape of the Gaussian is the same for the three groups (though obviously
their weight is different). Models with non-zero correlation coefficients
between the two variables are not favored in the BIC sense, contrary to the
models with $r=0$. 

\begin{figure}
{\includegraphics[width=.5\textwidth]{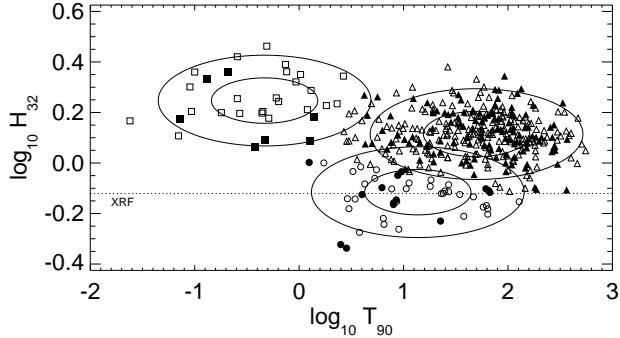}}
\caption{GRB populations on the duration-hardness plane. Different symbols
mark different groups. One and two sigma ellipses are superimposed on the
figure to illustrate the model components found as described in the text.
Filled symbols mark bursts with measured redshifts. The dashed line indicates
the definition of X-ray flashes (XRFs) given by \citet{2008ApJ...679..570S}. 
}
\label{3csop}
\end{figure}

\subsection{Non-parametric clustering}
A major draw-back of the model based clustering is that it assumes the
distribution of the underlying populations is of a given { functional} form
(bivariate Gaussian in our case). Non-parametric clustering does not assume any
{\it a priori} model. We need to define a metric to measure distances on the
duration-hardness plane.  Here we scale the variables with their standard
deviations because the clustering algorithms are sensitive to the distance
scale of the variables. If one of the variables has a standard deviation, for
example, one order of magnitude larger than the other, the method will use that
variable with greater weight.  Non-parametric clustering gives {\it definite}
membership values for each burst without providing any information on the
uncertainties of the clustering.  Here we perform k-means and hierarchical
clustering to substantiate our findings with the model based method.

\subsubsection{K-means clustering}
We apply k-means clustering to the dataset \citep[for an application of this
method see e.g.,][]{chat07}.  When applying this method we must know in advance
the number of clusters. Once the number of groups is known we find the
corresponding number of centers which minimizes the sum of squares to the
center of the group to which they belong. This is an iterative procedure.
There is no certain way of telling the "good" number of clusters. A speculative
method is to plot the within-group sum of squares as a function of the number
of clusters and look for "elbows" \citep{hartigan}. This would indicate that by
adding an extra group to the current number of groups, the explained variance
has fallen by a smaller amount than before, signalling that the addition of an
extra component is unnecessary.  \cite{1993A&A...268..108P} used the Akaike
Information Criterion to find the number of classes using k-means
clustering.  

\begin{figure} 
{\includegraphics[angle=-90,width=.5\textwidth]{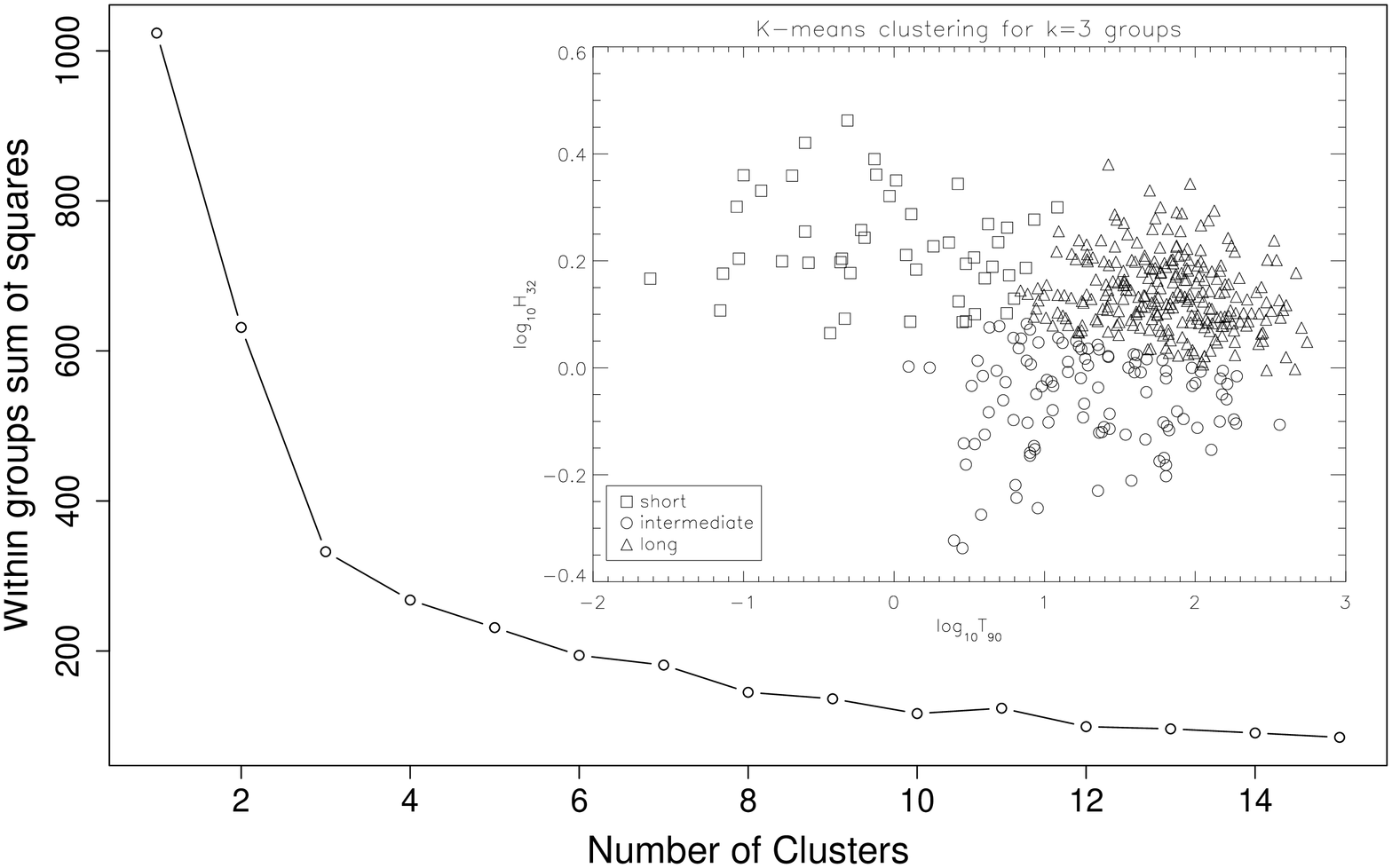}}
\caption{The evolution of the sum of squares while increasing the number of
groups in k-means clustering. An "elbow" is clearly visible at $k=3$. The inset
shows the group structure for $k=3$ groups on the duration-hardness plane.}
\label{kmplot} \end{figure}

We find a clear "elbow" feature on the number-of-clusters vs. within-groups-sum
of squares plot (Fig. \ref{kmplot}). Hence we deduce that, according to the
k-means clustering method, the most probable number of clusters is $k=3$. The
number of bursts in each group for $k=3$ as well as the center of the groups is
shown in Table \ref{km1}. This result strongly supports the group structure
found with the model-based method. 

\begin{table}  
\begin{center}
\caption{Group structure properties using k-means clustering for three populations. \bigskip}
\label{km1}
  \begin{tabular}{lccc} \hline
  Group				& N (\%)  		& Center($ T_{90}$ [s])& Center($H_{32}$)  \\
 \hline \hline
short             		&  $48$ ($11.8$)   	& $0.96$   	& $1.68$ 	\\
intermediate    		&  $105$ ($25.7$)  	& $20.1$   	& $0.87$  \\
long			      	&  $255$ ($62.5$)  	& $65.6$   	& $1.37$ 	\\
\hline
\end{tabular}
\end{center}
\end{table}

\subsubsection{Hierarchical clustering}
Another method of classifying bursts is the hierarchical clustering algorithm
\citep{1987ASSL..131.....M}.  We start from a state, where there are $N$ groups
(each burst is a separate group) and step by step we merge two groups using
some pre-defined criterion.  In $N-1$ steps we end up with just one group (all
bursts belong to one single class). 

We need to make a choice for the distance measure between two points. We choose
the simple Euclidean distance. This choice is motivated by the small
correlation between the two variables (correlation coefficient $r=-0.12$). In
case of a stronger correlation  the Mahalanobis distance measure is recommended
\citep{mahalanobis1936}.  One needs to define a method how groups will be
merged through the aforementioned steps. We chose the average linkage method.
This defines distance between two groups as the average of all the distances
between the pairs of points chosen from the two groups.  

\begin{figure}
{\includegraphics[angle=0,width=.5\textwidth]{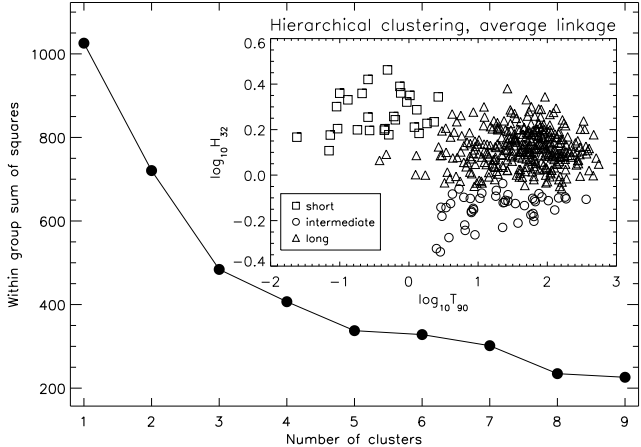}}
\caption{The within group sum of squares in function of the number of groups in
the case of the hierarchical classification.  Again, an "elbow" is visible at
$k=3$. The inset shows bursts classified
with hierarchical clustering. The structure of the
groups is similar to the model-based classification.}
\label{hclust}
\end{figure}

When applying the hierarchical clustering method, one gets the structure seen
in the inset of Fig. \ref{hclust} for $k=3$ groups. This resembles the group
structure found with model based clustering. As a justification for $k=3$
groups, we plot the within group sum of squares in function of the number of
groups as in the case of the k-means clustering and also see an "elbow" feature
at $k=3$. We thus conclude that three groups describe the sample
satisfactorily.

\subsection{Robustness of the clustering}\label{robust}
Using both model-based and non-parametric methods we have experimented by using
$T_{50}$ instead of $T_{90}$, by using different hardness ratios (e.g., $H_{42} =
\frac{S_{100-150}}{S_{25-50}}$, $H_{432} =
\frac{S_{100-150}}{S_{25-50}+S_{50-100}}$ etc.). The classification remained
essentially the same. 

The most "stable" group is the shortest and hardest population. The elements of
this group are clearly different from the other two and the membership remains
the same as we use other variants of hardness or duration.  The rest of the
bursts is divided between the long and the intermediate classes. At the border
between the classes bursts have a high class-uncertainty. This results in a
slight change of the membership of these bursts.  In other words, the separation
of the long and the intermediate population is fuzzy.

A classification is well founded if different methods give similar results. To
compare the similarities and differences between the hierarchical and the model
based classification { (EEI model)} we construct a so called contingency
table. This shows the number of bursts classified in the same- and different
groups by the two methods.  Table \ref{contingency}. shows that hierarchical
and model based classification schemes are consistent. The off-diagonal or
miss-classified elements ratio is only $4.4\%$.

The same table is made for the comparison of the k-means clustering and the
model-based clustering (see Table \ref{contingencykm}). The ratio of the
off-diagonal elements is higher in this case ($18.6 \%$). This is mainly caused
by the considerable overlap between the long and the intermediate groups, as
mentioned above.

\begin{table}  
\begin{center}
\caption{Contingency table for the hierarchical (HC) and the model based clustering. \bigskip}
\label{contingency}
  \begin{tabular}{lccccc}
				&  						& \multicolumn{3}{c}{Model based}\\
	\hline
				&						& Short	& Intermediate	& Long  & Total\\
 \hline 
				&Short	             		&  28 	& 0  	& 0		& 28 \\
HC				&Intermediate    		&  0  	& 39 	& 8  	& 47 \\
				&Long			      	&  3  	& 7   	& 323 	& 333\\
\hline
				&Total			      	&  31 	& 46  	& 331 	& 408\\
\hline
\end{tabular}
\end{center}
\end{table}

\begin{table}  
\begin{center}
\caption{Contingency table for the k-means (KM) and the model based clustering.\bigskip }
\label{contingencykm}
  \begin{tabular}{lccccc}
				&  						& \multicolumn{3}{c}{Model based}\\
	\hline
				&						& Short	& Intermediate	& Long  & Total\\
 \hline 
				&Short	             	&  31 	& 0  	& 17	& 48\\
k-means			&Intermediate    		&  0  	& 46 	& 59 	& 105\\
				&Long			      	&  0  	& 0   	& 255 	& 255\\
				\hline
				&Total		      		&  31 	& 46  	& 331 	& 408\\
\hline
\end{tabular}
\end{center}
\end{table}


\section{Peak-flux distribution}
Peak-flux is measured by Swift in the one-second interval about the highest
peak in the lightcurve. Counts are summed from this interval in the $15-150$
keV range in $58$ energy channels and deconvolved with the instrument's
spectral response matrix via a forward-folding method.  From the spectrum one
can obtain the peak-flux by integrating the best spectral model in the $15-150$
keV interval. The peak-flux is measured in units of ergs cm$^{-2}$ s$^{-1}$.

It is important to analyze if the intermediate population is in any ways
different from the other two.  In the previous section we have analyzed with
different methods the number of classes of bursts and determined the individual
burst's group membership.  After classifying the bursts on the
duration-hardness plane we compared the peak-flux distribution of the three
classes using Kolmogorov-Smirnov test.  In the following we use the classes
obtained by the EEI model-based classification.  

\begin{figure}
{\includegraphics[width=.5\textwidth]{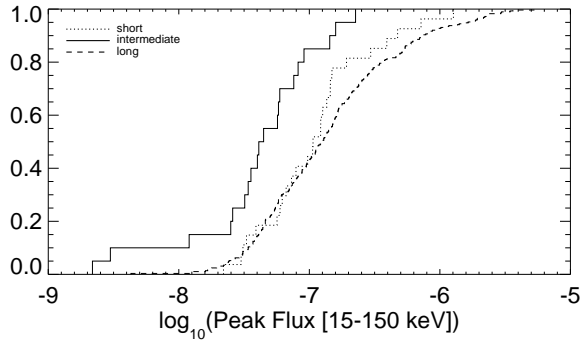}}
\label{petcumul}
\caption{Peak-flux cumulative distribution for the different groups. The
distribution of the short and long population is not significantly different. The
distribution of peak-fluxes of the intermediate class (dotted curve) is 
significantly different from the short ($0.7\%$) and long ($0.06\%$) groups.  }
\end{figure}

\begin{table}  
\begin{center}
\caption{Comparison of the three subgroups' peak-flux in the $15-150$ keV
range. This shows that the peak-flux distribution of the intermediate
population differs significantly from the long and the short population.\bigskip}
\label{tab2}
  \begin{tabular}{ccc} \hline
  Groups & KS distance & Error prob.\\
 \hline \hline
short-long             &   $0.221$ &    $0.167$\\
short-intermediate     &   $0.478$ &    $0.007$\\
long-intermediate      &   $0.448$ &    $6 \times 10^{-4}$\\
\hline
\end{tabular}
\end{center}
\end{table}

We found that the intermediate group has a different peak-flux distribution
with high significance ($ 6 \times 10^{-4}$; see Table \ref{tab2}) when
compared to the long population. In other words bursts which belong to this
group tend to have a lower peak-flux than both the long and the short
population.  It is worth mentioning that the other two { non-parametric}
classification methods led to similar conclusions. 

{
We thus found a difference in the peak-flux of the intermediate group of bursts.
It is also possible to include the peak-flux in the classification scheme.
Including the peak-flux in the classification, the main difference will be 
that the long duration group will be split in two, while the short and
intermediate groups will have the essentially the same members. 
}


\section{Populations with- and without measured redshift}
We have included the available redshift measurements for the  bursts.  The
distribution of each class was inspected for potential differences between the
groups.  We found that $23$ \% of bursts classified as short have measured
redshift ($7$ out of $31$). The ratio is slightly higher ($30 \%$ ($14$ out of
$46$) and, $36 \%$ ($119$ out of $331$)) for the intermediate and long
population respectively (for bursts with redshift see Fig. \ref{3csop}).

\begin{figure}
{\includegraphics[width=.5\textwidth]{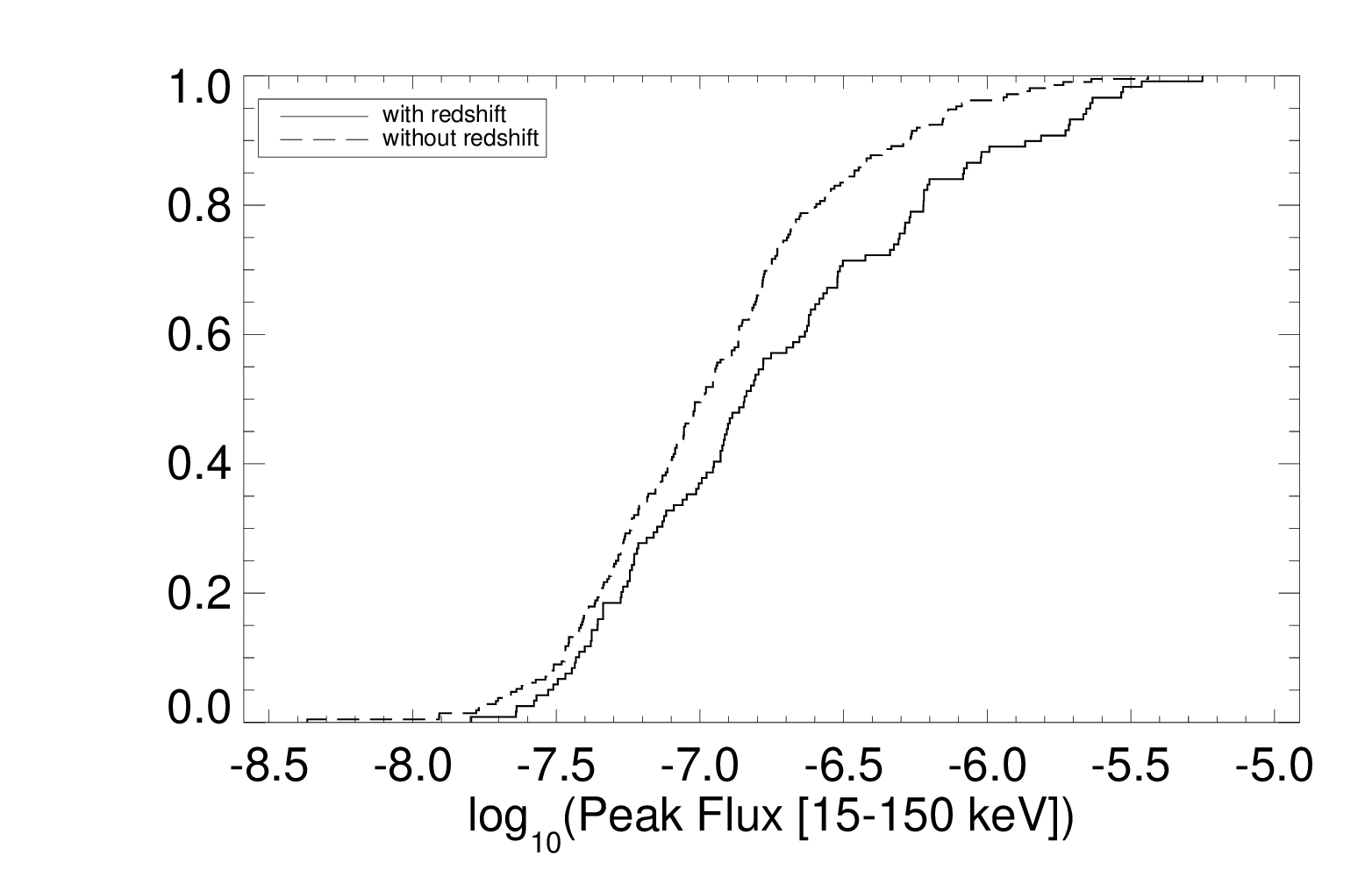}}\label{petotlong}
\caption{Cumulative distribution of the long population with- and without
measured redshifts. Long bursts with redshift have a clearly higher peak-flux
distribution.} 
\end{figure}

We have analyzed the distribution of the peak-flux of bursts in different
groups comparing the bursts with and without measured redshift. We found that
the peak-flux distribution of the long class with redshift measurement is
significantly different from the population without it.  Bursts with redshift
tend to have higher peak-flux than bursts without redshift (see Fig.
\ref{petotlong}). In other words, bursts with higher peak-flux have a better
chance of having a redshift measured.  There is no significant difference
between the other populations (see Table \ref{tab3}). 

\begin{table}  
\begin{center}
\caption{Comparison of the three subgroups' peak-flux in the $15-150$ keV
range. Here we compare peak-fluxes of different populations with and without
redshifts. We find there is a significant difference in the peak-flux
distribution of long bursts with measured redshift and the bursts
without.\bigskip}
\label{tab3}
\begin{tabular}{ccc} \hline
Sample    							& KS dist. 	& Error prob.\\
\hline \hline
Short z and non-z sample            	&     $0.208$ 	&      $0.955$\\
Interm. z and non-z sample     		&     $0.250$ 	&      $0.581$\\
Long      z and non-z sample   		&     $0.183$ 	&      $0.014$\\
\hline
\end{tabular}
\end{center}
\end{table}

\begin{figure}
{\includegraphics[width=.5\textwidth]{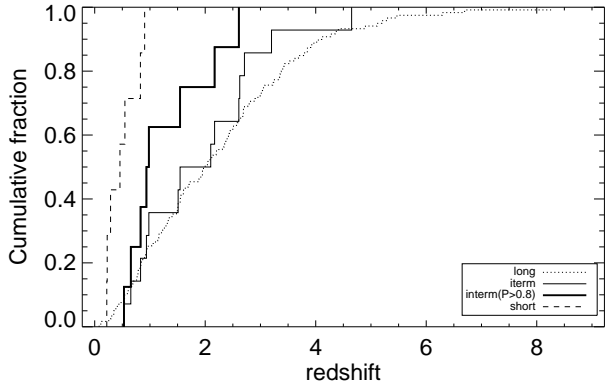}}\label{zdistr}
\caption{Cumulative redshift distribution of the three groups. As previously
known, short bursts are on average closer than the long GRBs. The intermediate
population distribution hints lower $z$ values than the long class, but the two
distributions are still compatible with the hypothesis of being drawn from the
same parent distribution. If we truncate the probability of the intermediate
class at $80 \%$, we find the difference is more apparent, but still not
significant. } \end{figure}

Next, we compare the distribution of the redshifts for the three groups with
each other. It is well-known, that short and long bursts differ significantly
in their redshifts \citep{bag06} and Swift bursts have a larger mean redshift
than previous spacecrafts' sample \citep{2006A&A...447..897J}.  We also find
that the distribution of the short bursts is markedly different when compared
to the long or the intermediate class (the error probability is $0.002$ and
$0.008$, resp.) (see Fig. \ref{zdistr}). The intermediate group has the same
redshift distribution as the long group (error probability: $\sim 0.79$).

The average redshift of the intermediate population is lower than the redshift
of the long group, but the difference is not significant. As we mentioned
earlier each burst is assigned to a particular group using the indicator
function.  Each burst has a finite probability of belonging to any of the three
groups. We assign the group membership of each burst based on the highest
probability between the three groups. We can restrict this criterion requiring
minimum value, e.g., $80 \%$ of the indicator function for a burst to belong to
a given group. Thus we will have less bursts in groups but more confident group
memberships. We have investigated the redshift distribution using this cut and
found that the long and intermediate groups  { seem} to be more different
(intermediate bursts are closer) but owing to the small number of bursts, this
difference is not significant (see Fig. \ref{zdistr}, the error probability has
decreased from $0.79$ to $0.19$).  

\section{Discussion}
\subsection{Physical interpretation}

Our analysis on the Swift GRBs supports the earlier results that there are
three distinct groups of bursts.  Again, besides the long and the short
population, the intermediate duration class appears to be the softest.  { In
this study, however, the structure of the intermediate class is not exactly the
same as in previous studies, mostly due to the different mathematical approach.
Due to the differences, it is possible to give a different physical
interpretation of the intermediate group, compared to previous studies, for
example linking them to X-ray flashes.  
}

A physical relationship of the intermediate class with the short population is
unlikely. This is suggested by almost non-existent contamination of short
bursts with intermediate in the cross-tabulated values with both hierarchical
and k-means clustering versus model based clustering.

The different { model based} classification algorithms reveal a significant
overlap between the distribution of the intermediate and the long class in the
duration-hardness plane.  One possibility  is that the intermediate class is a
distinct class by its physical nature. This may indicate there is  a third type
of progenitor.  

Also it is possible that the intermediate bursts do not form a different class
by themselves, but are related to the long population through some physically
meaningful parameter { or parameters}.  This could be the observing angle to
the jet axis \citep{2007ChJAA...7....1Z}, a less energetic central engine
possibly related through the angular momentum of the central black hole and the
accretion rate \citep{2010LNP...794..265K}, a baryon-loaded jet with lower
Lorentz factor \citep{1999ApJ...513..656D} or a combination of these.  This way
the intermediate population { represents a continuation of} the long
population.

The intermediate bursts' peak-flux are systematically lower than the long ones,
while their redshift range is either lower or similar.  We thus conclude that
the intermediate class is  intrinsically dimmer.  If the intermediate
population is part of the long population, the lower peak-flux requires a
physical explanation.  The observational properties show that intermediate
bursts are the softest among the three groups, meaning that their emission is
concentrated to low-energy bands.

\subsection{Relation to X-ray flashes}
As the intermediate population is the softest, it is worth searching for a link
with the similar and softer phenomenon compared to classical gamma-ray bursts,
the X-ray flashes (XRFs) (for a review, see \citet{2006PhDT.........5H}).
\citet{2008ApJ...679..570S} gives a working definition for X-ray flashes (XRF)
and X-ray rich GRBs (XRR) for Swift using the fluence ratio. The $S_{23}$
fluence ratio is the reciprocal of the hardness ($H_{32}=(S_{23})^{-1}$).
Current understanding of XRFs indicate that they are related to long bursts and
they form a continuous distribution in the peak energy ($E_{\rm{peak}}$) of the
$\nu F_{\nu}$ spectrum \citep{2008ApJ...679..570S}.

{
X-ray Flashes were first defined using BeppoSAX \citep{2001grba.conf...16H}. 
The criteria for an X-ray Flash was to trigger the Wide Field Camera (sensitive
between $2-30$ keV) instrument but not in the GRBM (sensitive between $40-700$
keV). 9 out of 10 XRFs detected by BeppoSAX were found in the BATSE data as
untriggered events \citep{2003AIPC..662..244K} with their bulk properties
similar to GRBs.
}

The  clustering methods  identifies on the duration-hardness plane the location
of the bursts in the intermediate class. According to the { fuzzy
classification} model we do not get a definite membership for a given burst,
rather a probability that a burst belongs to a group. To identify the
intermediate population (and tentatively the X-ray flashes), we use the
indicator function:

\begin{widetext}
\begin{eqnarray}\label{pxrf}
I_{\mathrm{Interm.}}(\log_{10} T_{90},\log_{10} H_{32})
=\frac{P_\mathrm{interm.}\times P(\log_{10} T_{90},\log_{10}
H_{32}|"Interm.")}{\displaystyle\sum_{l
\in \mathrm{\{short, interm., long\}}} P_\mathrm{l}\times P(\log_{10}
T_{90},\log_{10} H_{32}|l)}
\end{eqnarray}
\end{widetext}

\begin{figure}
{\includegraphics[width=.5\textwidth]{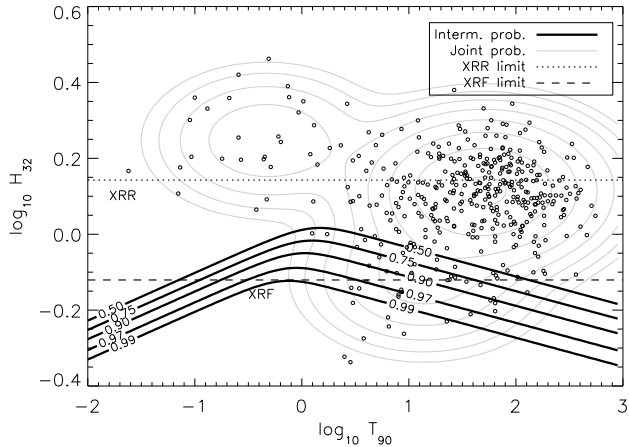}}
\caption{Contour plot of the duration-hardness distribution based on the EEI
model with three components in light-gray. Points show individual bursts. The
broken lines in black show the probability contours of a given region belonging
to the intermediate population. Also bursts classified as XRFs and XRR GRBs are
marked on the plot with horizontal lines. One can observe a remarkable
coincidence between the XRFs and the third group as shown by the indicator
function.}
\label{cplot}
\end{figure}

The values of the parameters in this equation should be taken from Table
\ref{modelparam}.  This yields the probability that a burst belongs to the
third group given its two measured parameters.  The joint distribution function
of the fitted model can be seen in gray on Fig. \ref{cplot} and the probability
contours of the third population are drawn in black with probability level
contours shown. We have also plotted the borders in hardness for the working
definitions of XRRs and XRFs in Fig. \ref{cplot} (dotted and dashed horizontal
lines respectively).

\cite{2008ApJ...679..570S} define XRFs as events with fluence ratio $S_{23} >
1.32$.  This translates to a hardness ratio $H_{32} < 0.76$. { This definition
aims to transform the limit of XRFs and X-ray rich GRBs found with BeppoSAX and
HeteII. The limit is found using a pseudo-burst with spectral parameters:
$\alpha =-1, \beta=-2.5 $ and $E_\mathrm{peak}=100$ keV for a Band spectrum
\citep{1993ApJ...413..281B}.} Based on this definition we identify $24$ bursts
from our $408$ burst sample. Table \ref{xrftab1}. contains data of bursts along
with the probabilities that they belong to the third population.  The average
of these probabilities (i.e. the XRF belongs to the intermediate group) is
$95\%$. This high value allows us to conclude that all XRFs belong to the
intermediate group { defined by the EEI model} with high probability. 

\begin{table}  
\begin{center}
\caption{X-ray flashes as defined by \cite{2008ApJ...679..570S} and the
probability that they belong to the intermediate group. \bigskip}
\label{xrftab1}
  \begin{tabular}{cccc} \hline
Name 	& $T_{90}[s]$  & $H_{32}$	& 3rd group prob.\\
\hline \hline
050416A & $  2.50$ & $ 0.48$ & { $ 1.00$} \\
050714B & $ 46.70$ & $ 0.73$ & { $ 0.80$} \\
050815 & $  2.90$ & $ 0.72$ & { $ 0.98$} \\
050819 & $ 37.70$ & $ 0.62$ & { $ 0.98$} \\
050824 & $ 22.60$ & $ 0.59$ & { $ 0.99$} \\
051016B & $  4.00$ & $ 0.75$ & { $ 0.97$} \\
060219 & $ 62.10$ & $ 0.68$ & { $ 0.89$} \\
060428B & $ 57.90$ & $ 0.67$ & { $ 0.91$} \\
060512 & $  8.50$ & $ 0.71$ & { $ 0.96$} \\
060923B & $  8.60$ & $ 0.70$ & { $ 0.97$} \\
060926 & $  8.00$ & $ 0.68$ & { $ 0.98$} \\
061218 & $  6.50$ & $ 0.57$ & { $ 1.00$} \\
070224 & $ 34.50$ & $ 0.75$ & { $ 0.80$} \\
070330 & $  9.00$ & $ 0.55$ & { $ 1.00$} \\
070714A & $  3.00$ & $ 0.66$ & { $ 0.99$} \\
070721A & $  3.78$ & $ 0.53$ & { $ 1.00$} \\
080218A & $ 23.00$ & $ 0.76$ & { $ 0.84$} \\
080218B & $  6.40$ & $ 0.60$ & { $ 1.00$} \\
080315 & $ 64.00$ & $ 0.66$ & { $ 0.92$} \\
080520 & $  2.84$ & $ 0.46$ & { $ 1.00$} \\
080822B & $ 64.00$ & $ 0.63$ & { $ 0.95$} \\
081007 & $  8.00$ & $ 0.69$ & { $ 0.98$} \\
081109B & $128.00$ & $ 0.70$ & { $ 0.72$} \\
081211A & $  3.44$ & $ 0.72$ & { $ 0.98$} \\
\hline
\end{tabular}
\end{center}
\end{table}

Based on Fig. \ref{cplot}. we propose that the members of the third component
are { probably the} X-ray flashes. Therefore, using the model based
classification method we can give { probabilistic} definition for the X-ray
flashes based on the duration-hardness distribution. This definition defines
$22$ additional bursts that belong to the intermediate population and hence to
the XRFs.

{ All the X-ray flashes are in the region where the third component has the
highest probability, but not all third component bursts can be unambiguously
classified as X-ray flashes according to the \citet{2008ApJ...679..570S}
criterion. In other words the third component in the EEI model contains all the
X-ray flashes and some additional, very soft bursts.}

\begin{figure}
{\includegraphics[width=.5\textwidth]{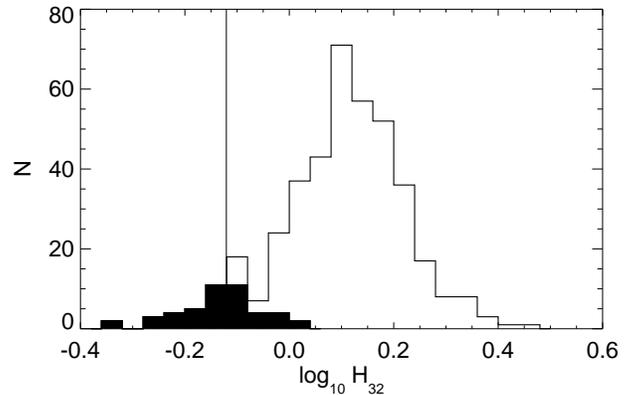}}
\caption{Hardness distribution of the bursts in the sample. The filled portion
marks the intermediate population. The vertical line shows the limit defined
to identify X-ray flashes by \cite{2008ApJ...679..570S}, it identifies $24$ XRFs.
An additional $22$ XRFs are proposed by the model fit probabilities.}
\label{histoxrf}
\end{figure}

To give further support to our point, we make a histogram with the hardness
ratios of the bursts (see Figure \ref{histoxrf}). The vertical line represents
the working definition of XRFs and the filled part of the histogram represents
putative X-ray flashes identified as the third, soft class in this study.  The
limiting contours are not horizontal, as the centers of the long and
intermediate classes have different $T_{90}$ values. Furthermore, there are
some short XRFs $T_{90}\approx 1 $s which are harder than the working
definition limit.

The mechanism behind the X-ray flashes is still not clear. There are various
scenarios that could produce these phenomena (e.g. dirty fireballs, inefficient
internal shocks, structured jets with off-axis viewing angle, etc., for a
review of the models see \citep{2007ChJAA...7....1Z}).  A more precise
experimental definition of XRFs can result in more stringent constraints on the
models.

\subsection{Relation to a recent study on groups of GRBs using Swift data} A
recent work by \citet{hor10} also confirmed the existence of the third class in
the Swift database.  \citet{hor10} used a maximum likelihood fit with EM
algorithm. { In their model they applied no restrictions for the parameters
of the ellipses. This can be related to the VVV model in the MClust package
with maximum number of degrees of freedom. In our case the VVV model is
disfavored because of the lower BIC value caused by the higher number of free
parameters. However, one model (VEI) with only marginally lower BIC value than
the best model has a similar structure as the one found by \citet{hor10}.  We
have constructed contingency tables comparing the common bursts in the
\citet{hor10} with the model based results.  (see Tables
\ref{contingencyhor10eei} and \ref{contingencyhor10vei} for the comparison with
the EEI and VEI models) 

The common sample with the \citet{hor10} study  consists of $324$ GRBs.
According to the contingency table (Table \ref{contingencyhor10eei}) there are
$31$ bursts which are classified as intermediate in both studies. The main
difference between the classifications can be seen in the the total number of
intermediate bursts: \citet{hor10} classify $86$ bursts as intermediates, and
this study finds only $32$ (according to the EEI model).

The other important question is the number of X-ray flashes in the two kinds of
classifications. Using the \citet{2008ApJ...679..570S} definition there are
$19$ X-ray flashes in the common sample. These belong without exception in the
intermediate class with the highest probability according to {\it both}
\citet{hor10} and this study.  In other words the $31$ bursts classified as
intermediate by both \citet{hor10} and this study, contains all the X-ray
flashes. \citet{hor10} classifies $55$ bursts in the intermediate class,
whereas here we classify them as long. Based on this we can state that the
model presented here is more efficient to identify the X-ray flashes with high
probability.  The ratio of X-ray flashes in the intermediate class is $59\%$ in
the EEI model and $22\%$ in the \citet{hor10} study.

The VEI model with a marginally lower BIC value has only $32$ off-diagonal
elements in the contingency table (see Table \ref{contingencyhor10vei}) when
compared to the \citet{hor10} study. The structure revealed by this model is
more similar to the one in \citet{hor10}.  In this model all the $19$ X-ray
flashes are also classified as intermediate. In this case the the number of
intermediate bursts is $118$ which means there are many bursts classified as
intermediate which are not X-ray flashes. The ratio of XRFs to intermediate
class members is $16\%$. }

{
\begin{table}  
\begin{center}
\caption{Contingency table comparing the $324$ common bursts from \citet{hor10}
and the EEI solution used in this study. There are $19$ X-ray flashes in this
sample all of which are classified as intermediate by both methods, i.e. they
are included in the intermediate-intermediate field with $31$ elements.}
\label{contingencyhor10eei}
  \begin{tabular}{lccccc}
				&  						& \multicolumn{3}{c}{\citet{hor10}} classification\\
	\hline
				&						& Short	& Intermediate	& Long  & Total\\
 \hline 
				&Short	             	&  24 	& 0  	& 0 	& 24\\
Model based     &Interm. 		   		&  0  	& 31	& 1 	& 32\\
(EEI)			&Long			      	&  0  	& 55  	& 213 	& 268\\
				\hline
				&Total		      		&  24 	& 86  	& 214 	& 324\\
\hline
\end{tabular}
\end{center}
\end{table}

\begin{table}  
\begin{center}
\caption{Contingency table comparing the $324$ common bursts from \citet{hor10}
and the VEI model.}
\label{contingencyhor10vei}
  \begin{tabular}{lccccc}
				&  						& \multicolumn{3}{c}{\citet{hor10}} classification\\
	\hline
				&						& Short	& Intermediate	& Long  & Total\\
 \hline 
				&Short	             	&  22 	& 0  	& 0 	& 22\\
Model based		&Interm.	    		&  2 	& 86	& 30	& 118\\
	(VEI)		&Long			      	&  0  	& 0  	& 184 	& 184\\
				\hline
				&Total		      		&  24 	& 86 	& 214 	& 324\\
\hline
\end{tabular}
\end{center}
\end{table}
}
{
The reason for finding different group structure for the intermediate
population lies in the fact that this group is { significantly} overlaid
with the long population and it is { very much} sensitive to the
mathematical approach used.  }

\section{Conclusion}
The results of this paper can be summarized as follows:
\begin{itemize}
\item[-] We have established with multiple methods - in concordance with
	previous studies - that Swift GRB data can be best modelled using three
	populations.  Both the model independent and the model based methods
	showed three groups with high significance.

\item[-] We found that the third population of GRBs, intermediate in duration
	and with the softest spectrum, has a peak-flux distribution that
	significantly differs from the other two classes. This group has the
	lowest average  peak-flux. 

\item[-] Furthermore, the redshifts of the intermediate population do not
	differ significantly from that of the long class, although their
	average redshift is lower.  Considering this and their lower average
	peak-flux it indicates that the intermediate GRBs are inherently dimmer
	than the longer ones.

\item[-] We have also found evidence that the intermediate population is
	closely related to X-ray flashes: all the previously identified Swift
	X-ray flashes belong to the third, soft population. Therefore, we
	{ relate} the intermediate class { to} the X-ray flashes.  Thus, we give
	a new, probabilistic definition for this phenomenon.  
\end{itemize}

\acknowledgements
This work was supported by OTKA grant K077795, by OTKA/NKTH A08-77719 and
A08-77815 grants (Z.B.), by the GA\v{C}R grant No. P209/10/0734 (A.M.), by the
Research Program MSM0021620860 of the Ministry of Education of the Czech
Republic (A.M.) and by a Bolyai Scholarship (I.H.).  We thank Peter
M\'esz\'aros, G\'abor Tusn\'ady, L\'idia Rejt\H{o}, Jakub \v{R}\'{\i}pa {
and the anonymous referee} for valuable comments on the paper.

\bibliography{lnls}

\end{document}